\documentclass[9pt]{article}
\usepackage{spconf,amsmath,graphicx}
\usepackage{amsfonts} 
\usepackage{mathtools}

\title{Cross-modal alignment with optimal transport for CTC-based ASR}
\name{Xugang Lu$^{1*}$, Peng Shen$^{1}$, Yu Tsao$^{2}$, Hisashi Kawai$^1$
	}
\address{1. National Institute of Information and Communications
	Technology, Japan.\\
	2. Research Center for Information Technology Innovation, Academic Sinica, Taiwan}
\begin{document}
%
\maketitle
\begin{abstract}
Temporal connectionist temporal classification (CTC)-based automatic speech recognition (ASR) is one of the most successful end to end (E2E) ASR frameworks. However, due to the token independence assumption in decoding, an external language model (LM) is required which destroys its fast parallel decoding property. Several studies have been proposed to transfer linguistic knowledge from a pretrained LM (PLM) to the CTC based ASR. Since the PLM is built from text while the acoustic model is trained with speech, a cross-modal alignment is required in order to transfer the context dependent linguistic knowledge from the PLM to acoustic encoding. In this study, we propose a novel cross-modal alignment algorithm based on optimal transport (OT). In the alignment process, a transport coupling matrix is obtained using OT, which is then utilized to transform a latent acoustic representation for matching the context-dependent linguistic features encoded by the PLM. Based on the alignment, the latent acoustic feature is forced to encode context dependent linguistic information. We integrate this latent acoustic feature to build conformer encoder-based CTC ASR system. On the AISHELL-1 data corpus, our system achieved 3.96 \% and 4.27 \% character error rate (CER) for dev and test sets, respectively, which corresponds to relative improvements of 28.39 \% and 29.42\% compared to the baseline conformer CTC ASR system without cross-modal knowledge transfer.                                   
 
\end{abstract}
\section{Introduction}
\label{sec:intro}
End to end (E2E) model has achieved substantial improvement in automatic speech recognition (ASR) in recent years with several state of the art model frameworks \cite{Li2022}, for example, temporal connectionist temporal classification (CTC)-based ASR \cite{CTCASR}, attention with encoder-decoder (AED)-based ASR \cite{Chan2016, Kim2017, Hori2017,Watanabe2017}, and recurrent neural network transducer (RNN-T)-based ASR \cite{RNNTASR}, etc. Among these E2E-ASR frameworks, CTC-based ASR attracts a lot of attention. One of its advantages is its non-autoregressive (NAR) decoding capability, i.e., fast and parallel decoding in obtaining transcription tokens since tokens in CTC are assumed to be independent. However, this token independence assumption in CTC makes it difficult for acoustic encoder to learn rich context dependent linguistic information during model training. Therefore, an external language model (LM) is often required as a post processing to improve the ASR performance. In recent years, the effectiveness of pretrained language models (PLMs) in natural language processing (NLP) tasks has led to their frequent usage as external language models for rescoring in ASR \cite{BERTScore,MLMScore}. Using an external LM as another post-processing (e.g., beam search, rescoring, etc.) destroys the fast and parallel decoding property of CTC-based ASR. So here comes a question: is it possible to explicitly encode rich linguistic information in acoustic feature representation for the CTC-based ASR without using any external LM for post processing. In this study, rather than utilizing a PLM as an external LM, we focus on how to transfer the context-dependent linguistic knowledge from a PLM to acoustic feature representation learning, and do recognition with CTC-based ASR only.

One of the most successful CTC-based E2E ASR frameworks which enhances linguistic information in acoustic encoder is based on a hybrid CTC/AED-based ASR model framework \cite{Chan2016,Kim2017, Hori2017,Watanabe2017}. In this framework, an attention text decoder with cross entropy induced attention loss and CTC based loss are integrated in a multi-task (or objective) learning framework. The integration of the text decoder in model learning allows the shared acoustic encoder to potentially learn rich linguistic information \cite{T5,SXLSR2022}. With multi-task learning framework, several methods have been proposed to learn linguistic information by inserting linguistic knowledge in intermediate layers of acoustic encoders for ASR \cite{HierarchicalCTC,intermediateCTC}. In recent years, due to the success of self-supervised learning in feature exploration, knowledge transfer learning from both pretrained acoustic model (e.g., wav2vec2.0 \cite{wav2vec2.0}) and PLM (e.g., bidirectional encoder representation from transformers (BERT) \cite{BERT}) for ASR task also have been proposed \cite{CIFBERT1,CTCBERT1,CTCBERT2,CIFBERT2,Cho2020,FNAR-BERT,wav2vecBERTSLT2022}. There are two strategies for transferring linguistic knowledge from a PLM to ASR, one strategy is to stack text encoder of the PLM on top of the acoustic encoder in an ASR framework and fine tune both encoders for the ASR task. The other strategy is to design a multi-task (or objective) learning framework to transfer linguistic knowledge from the text encoder branch to acoustic encoder branch for the ASR task. Several studies have been proposed based on these two strategies. For example, in \cite{NARBERT}, it was suggested to stack a BERT text encoder on a transformer-based acoustic encoder, allowing an NAR-based ASR model to leverage the pretrained BERT-based language model as a decoder for recognition. However, PLMs in NLP are often trained for common NLP tasks such as question answering, text summarization, and others. They are not compatible with ASR tasks because these LMs are designed to process standard text inputs. Moreover, stacking the PLM in ASR decoding increases the inference complexity. From the point view of knowledge distillation (KD) \cite{KD2015}, when transferring linguistic information encoded in a PLM to an acoustic model, the PLM can be considered as the teacher model while the acoustic model serves as the student model. The rich textual knowledge from the teacher model can be distilled to the student model through cross-modal knowledge distillation during model training \cite{Choi2022,Cross2021,Futami2022,Higuchi2023}. During recognition, only the student model is used, and the teacher model is not involved in the decoding process for ASR. 

Speech and text sequences, like two sides of a coin, possess shared knowledge that is represented in two modalities, but their representation distributions and lengths are quite different. 
Most of the methods mentioned above encounter a common problem, namely, how to efficiently align feature representations between text and acoustic modalities to facilitate transfer knowledge from a PLM to an acoustic encoder. One of the most successful alignment algorithms is based on a cross-modal attention (CMA) modeling between the acoustic and text representations within a transformer decoder framework \cite{Transformer}. However, when the CMA based decoder is not involved in decoding for ASR (e.g., the NAR based ASR trained with CTC only), the efficiency of linguistic information encoded in the acoustic encoder is limited. In this study, we propose a novel cross-modal alignment method for transferring linguistic information encoded in a PLM for ASR. Our method is based on optimal transport (OT) which was originally proposed in mathematics for measuring discrepancy between two probability distributions \cite{VillanoBook}. The OT has been applied for shape matching and domain adaptation in machine learning \cite{CourtyNIPS2017,ICLR2023}, in cross-domain spoken language recognition and speech enhancement \cite{LuICASSP2021,Lin2021}, in speech translation \cite{ACL2023, ICML2023}. In this study, we adopt the OT for cross-modal alignment and knowledge transfer for ASR. Our contributions can be summarized as follows: 

1. We propose a cross-modal alignment and knowledge transfer model based on OT and integrate the cross-modal transfer loss with the CTC-based loss for training the acoustic model. 

2. To enable efficient knowledge transfer, we introduce a cross-modal neural adapter that facilitates the transfer linguistic knowledge from a PLM to the acoustic encoder. 

3. We construct an ASR system using the proposed cross-modal transfer learning algorithm and validate its effectiveness through experiments.  

The remainder of this study is organized as follows: Section \ref{sec:proposed} presents the proposed model framework in which the cross-modal alignment based on OT is introduced. In Section \ref{sec:exp}, the experiments conducted are evaluated, and comparisons with several advanced knowledge transfer learning algorithms for ASR are presented. Finally, the conclusion is presented in Section \ref{sec:conclusion}.   

\section{Proposed method}
\label{sec:proposed}
\begin{figure}[tb]
	\centering
	\includegraphics[width=7cm, height=6cm]{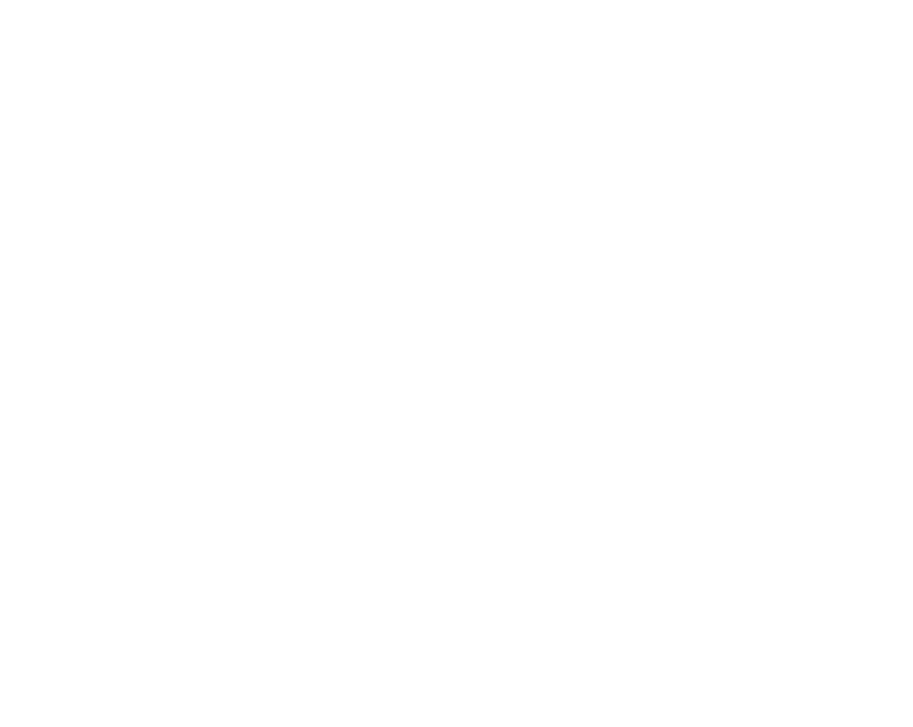}
	\caption{Cross-modal matching and knowledge transfer. Only modules or blocks in the dashed red box are used in inference (recognition).}
	\label{fig:framework}
\end{figure}

In this study, the two modalities are an acoustic encoder with CTC based acoustic model and BERT based text model. The purpose for cross-modal knowledge transfer in our study is to transfer the linguistic knowledge encoded in BERT to the acoustic encoder for ASR. For efficient knowledge transfer, an OT based alignment and transform is designed. The proposed model framework is showed in Fig. \ref{fig:framework}. In this figure, the two modalities are located in two branches, the left branch consists of encoder blocks and a full-connection layer (FC1) with softmax activation function which is trained with CTC-based loss for speech recognition. The right branch is for text processing based on a PLM (BERT). There is a connection between these two modalities via an OT matching block for cross-modal alignment. Moreover, a cross-modal neural adapter (the blocks in gray in the left branch): Two full-connection layers (FC2 and FC3) and two layer normalizations (LN) is attached to acoustic modality for efficient linguistic knowledge transfer. The final feature representation for ASR is an addition of the acoustic encoder output and cross-modal adapter output. It is supposed that the feature explored by this two-branch modalities will enhance the linguistic information representation for ASR. We will introduce the model framework in details in the following.
\subsection{Acoustic feature representation in speech modality}
In the left branch of Fig. \ref{fig:framework}, the original input acoustic feature is ${\bf X} \in {\mathbb{R}}^{T \times d} $, where $T$ is the time length, $d$ is the feature dimension. The `CNN blocks Subsampling' module is used for speech feature transform and subsampling, which is composed of two 2D convolution layers (with ReLU activation) and one feed-forward layer. And each convolution is with a stride larger than one for subsampling. After this subsampling process, the position encoding (PE) is added before it is input to the encoder blocks. The process is formulated as:
\begin{equation}
	\begin{array}{l}
		{\bf \tilde X} = {\rm CNNSubsampling}\left( {\bf X} \right) \\ 
		{\bf H}_0^{in}  = {\bf \tilde X} + {\rm PE}\left( {{\bf \tilde X}} \right) \\ 
	\end{array}
	\label{eq:Subsample}
\end{equation}

Acoustic encoder in Fig. \ref{fig:framework} is a conformer based encoder \cite{conformer2020}. Each conformer block is composed of four consecutive modules, i.e., feed-forward network (FFN) (FNN1) module, multi-head self attention (MHSA) module, convolution (Conv) module, and a second FFN (FFN2) module. The processing in each conformer block is formulated as:
\begin{equation}
	\begin{array}{l}
		\vspace{2mm}
		{\bf H}_l^1  = {\bf H}_{l - 1}^{in}  + \frac{1}{2}{\rm FFN1(}{\bf H}_{l - 1}^{in} {\rm )} \\ 
		\vspace{2mm}
		{\bf H}_l^{2}  = {\bf H}_l^1  + {\rm MHSA(}{\bf H}_l^1 {\rm )} \\ 
		\vspace{2mm}
		{\bf H}_l^{3} = {\bf H}_l^{2}  + {\rm Conv(}{\bf H}_l^{2} {\rm )} \\ 
		\vspace{2mm}
		{\bf H}_l^{4} = {{\bf H}_l^{3}  + \frac{1}{2}{\rm FFN2(}{\bf H}_l^{3} {\rm )}}  \\ 
		\vspace{2mm}
		{\bf \tilde H}_l^4  = {\rm LN}\left( {{\bf H}_l^4 } \right) \\
		\vspace{2mm}
		{\bf H}_{l}^{in}  ={\bf \tilde H}_l^4,  \\ 		
	\end{array}	
	\label{eq:conformer}
\end{equation}
where $l$ takes values from $1$ to $L$, with $L$ representing the total number of blocks, ${\bf H}_{0}^{in}$ is the output of the CNN subsampling process with position encoding defined in Eq. (\ref{eq:Subsample}). ${\bf \tilde H}_L^4  \in \mathbb{R}^{T_a  \times d_a }$ is the final output of the conformer encoder with length (temporal dimension) $T_a$, and feature dimension $d_a$. This acoustic feature is utilized for speech recognition, wherein a linear projection layer (FC1) is applied prior to using softmax to estimate the probability (${\bf P}$) of predicted tokens as:
\begin{equation}
	{\bf P} = {\rm Softmax}\left( {{\rm FC1}\left( {{\bf \tilde H}_L^4 } \right)} \right)
	\label{eq:softmax}
\end{equation}
Please note that in CTC-based model training, the token sequence is used the same as in the text modality (BERT tokenization) (refer to \ref{sect:textmodal}), and the symbols `$\bf{e}_{{\rm bos}}$' and `$\bf{e}_{{\rm eos}}$' represent the start and end of acoustic sequences.
\subsection{Linguistic feature representation in text modality}
\label{sect:textmodal}
In the right branch of Fig. \ref{fig:framework}, the context-dependent linguistic representation is explored from a pretrained BERT model. The process is formulated as: 
\begin{equation}
	\begin{array}{l}
		{\bf y}_{{\rm token}}  = {\rm Tokenizer}\left( {\bf y} \right) \\ 
		\vspace{2mm}
		{\bf Z}_0  = \left[ {{\rm CLS, }{\bf y}_{{\rm token}} ,{\rm SEP}} \right] \\ 
		\vspace{2mm}
		{\bf Z}_i  = {\rm BERT}_i \left( {{\bf Z}_{i - 1} } \right), \\ 		
	\end{array}
	\label{eq:bert}
\end{equation}
where `${\rm BERT}_{i}$' is the $i$-th transformer encoder layer of BERT model, $i$ takes values from $1$ to $M$, with $M$ representing the total number of BERT encoder layers. `$\rm Tokenizer$' is a process to convert standard text to word piece based tokens \cite{BERT}. Token symbols `CLS` and `SEP` represent the start and end of an input sequence. ${\bf Z}_M  \in \mathbb{R}^{T_t  \times d_t } $ is the final text representation which encodes context dependent linguistic information, $T_t$ denotes the sequence length, and $d_t$ represents feature dimension of text encoding representation. The outputs of acoustic and text encoders,  ${\bf \tilde H}_L^4  \in \mathbb{R}^{T_a  \times d_a }$ and ${\bf Z}_M  \in \mathbb{R}^{T_t  \times d_t } $, are with different dimensional modalities, a cross-modal alignment is required before comparing their features. 	
\subsection{Cross-modal alignment based on OT matching}
The original OT concept is used to transport from one probability distribution to another with minimum amount of transport cost \cite{VillanoBook}. With a relaxed usage of the concept, we can define OT for cross-modal alignment when we regard the acoustic and text feature sequences as two independent distributions \cite{Chi2021}. By finding an optimal transport, a latent acoustic feature can be transformed to a space which is guided by the context dependent linguistic information. We define the cross-modal OT as:
\begin{equation}
	L_{{\rm OT}} ({\bf Z},{\bf H})\mathop  = \limits^\Delta  \mathop {\min }\limits_{\gamma  \in \prod {\left( {{\bf Z},{\bf H}} \right)} } \sum\limits_{i,j} {\gamma \left( {{\bf z}_i ,{\bf h}_j } \right)C\left( {{\bf z}_i ,{\bf h}_j } \right)}, 
	\label{eq:OT}
\end{equation} 
where $\gamma  \in \mathbb{R}^{T_t  \times T_a }$ is a transport plan matrix with dimensions $T_t$ and $T_a$, ${\prod {\left( {{\bf Z},{\bf H}} \right)} }$ is the set of transport plan between two distributions of text feature $\bf Z$ and acoustic feature $\bf H$. $C\left( {{\bf z}_i ,{\bf h}_j } \right)$ is a transport cost or distance function between ${\bf z}_i$ and ${\bf h}_j$, where ${\bf z}_i$ is the $i$-th column vector of $\bf Z$, and ${\bf h}_j$ is the $j$-th column vector of $\bf H$. 

In Eq. (\ref{eq:OT}), for matching the feature dimension to that of text representation, the acoustic latent feature $\bf H$ is defined as a linear transformation of the conformer encoder output as:
\begin{equation}
	{\bf H} = {\rm FC2}\left( {{\bf \tilde H}_L^4 } \right) \in \mathbb{R}^{T_a  \times d_{t} },
\end{equation}
where $\rm FC2(.)$ is a linear transform function. And the text feature $\bf Z$ is the output of the BERT encoder, i.e., ${\bf Z} = {\bf Z}_M  \in \mathbb{R}^{T_t  \times d_t } $. Cost function ${C\left( {{\bf z}_i ,{\bf h}_j } \right)}$ is defined as a cosine distance:
\begin{equation}
	C\left( {{\bf z}_i ,{\bf h}_j } \right) = 1 - \cos \left( {{\bf z}_i ,{\bf h}_j } \right)
	\label{eq:cosine}
\end{equation} 
For efficiently solving Eq. (\ref{eq:OT}), we seek to solve an entropy regularized OT (EOT) defined as \cite{OPT}:
\begin{equation}
	L_{{\rm EOT}} ({\bf Z},{\bf H})\mathop  = \limits^\Delta  \mathop {\min }\limits_{{\gamma  \in \prod {U( {{\bf Z},{\bf H}})}}  } {\sum\limits_{i,j} {\gamma ( {{\bf z}_i ,{\bf h}_j } )C( {{\bf z}_i ,{\bf h}_j } )} - \alpha H(\gamma )}, 
	\label{eq:EOT}
\end{equation}
where $H(.)$ is the entropy function of transport coupling, and $\alpha >0$ is a regularization coefficient. The solution can be obtained as:
\begin{equation}		
	L_{{\rm EOT}} ({\bf Z},{\bf H}) = \sum\limits_{i,j} {\gamma ^* \left( {{\bf z}_i ,{\bf h}_j } \right)C\left( {{\bf z}_i ,{\bf h}_j } \right)}  - \alpha H(\gamma ^* ), 			
	\label{eq:OTsolve}
\end{equation}
where the optimal coupling $\gamma ^*$ is estimated by:
\begin{equation}
	\gamma ^* \mathop  = \limits^\Delta  \mathop {\arg \min }\limits_{\gamma  \in \prod {\left( {{\bf H},{\bf Z}} \right)} } L_{{\rm EOT}} ({\bf Z},{\bf H}) 
	\label{eq:coupling}
\end{equation}
\subsection{Cross-modal knowledge transfer}
\begin{figure}[tb]
	\centering
	\includegraphics[width=8cm, height=4cm]{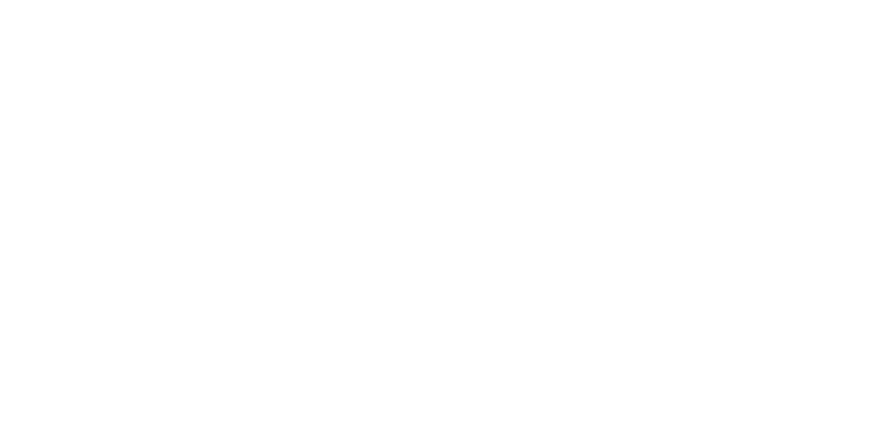}
	\caption{Linguistic guided acoustic representation with OT based matching}
	\label{fig:optloss}
\end{figure}

After obtaining the optimal coupling from Eq. (\ref{eq:coupling}), the context dependent linguistic guided acoustic representation is estimated as:
\begin{equation}
	{\bf \tilde Z}_{{\bf Z} \leftarrow {\bf H}} \mathop  = \limits^\Delta  \gamma ^*  \times {\bf H} \in \mathbb{R}^{T_t  \times d_t }  
\end{equation}
We summarize the transform for context dependent linguistic guided acoustic feature extraction in Fig. \ref{fig:optloss}. 
From this figure, we can observe that a bi-level optimization framework involves OT matching and feature transformation. The OT optimization is embedded within the optimization process for cross-modal feature transformation. Through this transform, the representation from acoustic modality space is projected onto the text modality space, allowing for the direct estimation of cross-modal discrepancy (or cross-modal alignment loss) by:
\begin{equation}
	L_{{\rm align}}  = \sum\limits_{i = 2}^{T_t  - 1} {1 - \cos \left( {{\bf z}_i ,{\bf \tilde z}_i } \right)}
	\label{eq:align}
\end{equation}
In this formulation, the sum ranges from 2 to $T_t-1$ in order to exclude the `[CLS]' and `[SEP]' tokens from the loss estimation (refer to Eq. (\ref{eq:bert}) in text encoding). 

By minimizing the alignment loss defined in Eq. (\ref{eq:align}), it is supposed that the feature representation $\bf H$ is pushed to encode rich linguistic information. For making this representation compatible with acoustic feature for ASR, the following transforms are designed:
\begin{equation}
	\begin{array}{l}
		{\bf \hat H} = {\rm FC3}\left( {{\rm LN}\left( {\bf H} \right)} \right) \in R^{T_a  \times d_a }  \\ 
		{\bf H}_{a,t}  = {\bf \tilde H}_L^4  + s \cdot {\rm LN}\left( {{\bf \hat H}} \right) \\ 
	\end{array}
	\label{eq:adapter}
\end{equation}
In Eq. (\ref{eq:adapter}), `FC3' is a full-connected linear transform, `LN' is a layer normalization operator, $s$ is a weighting coefficient. Based on this new representation ${\bf H}_{a,t}$ (subscript `$_{a,t}$' denotes features incorporating acoustic and text modalities), the probability prediction for recognition is:
\begin{equation}
	{\bf \tilde P} = {\rm Softmax}\left( {{\rm FC1}\left( {{\bf H}_{a,t} } \right)} \right)
	\label{eq:softmaxadd}
\end{equation}

In training for cross-modal knowledge transfer, given an input acoustic feature sequence and corresponding output text token sequence, the total loss is defined as:
\begin{equation}
	L \mathop  = \limits^\Delta  \lambda  \cdot L_{{\rm CTC}} ({\bf \tilde P},{\bf y}_{{\rm token}} ) + \left( {1 - \lambda } \right) \cdot w \cdot (L_{{\rm align}}  + L_{{\rm EOT}} )  
	\label{eq:totalloss} 
\end{equation}
where $L_{{\rm CTC}} ({\bf \tilde P},{\bf y}_{{\rm token}} )$ is CTC loss, $L_{{\rm align}}$ is cross-modal alignment loss defined in Eq. (\ref{eq:align}), and $L_{{\rm EOT}}$ is OT loss defined in Eq. (\ref{eq:OTsolve}), $\lambda$ is a trade off parameter, $w$ is a parameter to scale the alignment loss. After the model is trained, only the left branch of Fig. \ref{fig:framework} is kept for ASR inference.  
\section{Experiments}
\label{sec:exp}
In this section, we evaluate our proposed cross-modal alignment and knowledge transfer model on the ASR task to determine whether the new acoustic representation guided by the linguistic features from the PLM can improve ASR performance or not. 
\subsection{Data corpus and feature process}
Our experiments are carried out on an open source Mandarin speech corpus AISHELL-1 which includes speech recorded from 400 speakers \cite{AISHELL1}. In our experiments, the corpus was divided into three sets, i.e., a training set with 340 speakers (150 hours), a development (or validation) set with 40 speakers (10 hours), and a test set with 20 speakers (5 hours). In training, we applied speed perturbation with factors of 0.9 and 1.1 for data augmentation the same as used in the baseline system for AISHELL-1 corpus \cite{AISHELL1}. The input feature vector for the deep model network is composed of two components, i.e., 80-dimensional log Mel-filter bank features, and 3-dimensional fundamental frequency related features (F0, delta F0 and delta delta F0). These features are computed with a 25ms window size and a 10ms shift. 
\subsection{Model implementation}
As showed in Fig. \ref{fig:framework}, before the speech features are sent to the acoustic encoder, a CNN transform and subsampling process module is applied. This CNN subsampling module is consisted of two 2D convolutional blocks (256 channels, kernel size 3, and stride 2, with ReLU activation function for each). In the acoustic encoder blocks, $L=16$ conformer blocks are used and each is with a series of computations defined in Eq. (\ref{eq:conformer}). Moreover, in the conformer block, convolutional kernel size is 15, attention dimension is 256, attention head is 4, and the dimension of FFN is 2048. The dimension of the feature output of the conformer encoder is $d_a=256$. In text processing blocks with Eq. (\ref{eq:bert}), the `bert-base-chinese' from huggingface is used in the BERT model \cite{Huggingface}. In this BERT model, there are $M=12$ transformer encoders, token size is 21128, and the dimension of the final text feature representation is $d_t=768$. In order to match feature dimensions between acoustic and text modalities, the linear transform `FC2' in Fig. \ref{fig:framework} is with size of $768*256$ weight matrix (to transform from acoustic feature with dimension $d_a=256$ to text feature with dimension $d_t=768$), and `FC3' is used to transform back (with matrix size of $256*768$) to the feature size of acoustic space. In AISHELL-1, there are only 4230 unique characters. By using Chinese BERT tokenizer model on AISHELL-1 text data, token sequences are obtained for supervised model training, and only 4281 unique tokens appeared among the 21128 token-based vocabulary size of Chinese BERT model. Based on this information, `FC1' used for class probability prediction in Eqs. (\ref{eq:softmax}) and (\ref{eq:softmaxadd}) is a linear transform to convert an acoustic feature vector to class prediction probability logits with length of 4282 (one additional token `BLK' used in CTC). Several hyper-parameters are used in the model, in this study we fixed their values during experiments as: EOT regularization parameter $\alpha=0.2$ in Eq. (\ref{eq:OTsolve}), weighting coefficient $s=1.0$ in Eq. (\ref{eq:adapter}), scale parameter $w=1.0$ and alignment trade off parameter $\lambda=0.3$ in Eq. (\ref{eq:totalloss}). 

Adam optimizer \cite{Adam} is used with a learning rate (initial with 0.001) schedule with 20,000 warm-up steps. The model was trained for 130 epochs, and the final model used for evaluation was obtained by averaging models from the last 10 epochs. The performance was evaluated based on character error rate (CER).   
\subsection{Results}
\label{sec:results}
After the model is trained, only the left branch (blocks in dashed red box in Fig. \ref{fig:framework}) is used for ASR, i.e., CTC based decoder is used in speech recognition. In addition, we only used CTC greedy search in decoding, the results are showed in table \ref{tab1}. The results of baseline system and several state-of-the-art systems which integrate BERT for linguistic knowledge transfer are also showed in Table \ref{tab1} for comparison. 
\begin{table}[tb]
	\centering
	\caption{ASR performance on AISHELL-1 coprus, CER (\%).}
	\begin{tabular}{|c||c||c|}
		\hline
		Methods &dev set &test set\\
		\hline
		Conformer+CTC (Baseline)  &5.53 &6.05 \\
		\hline	
		\hline	
		Conformer+CTC/AED (\cite{Watanabe2017,wenet2.0})  &4.61 &5.06 \\						
		\hline
		NAR-BERT-ASR (\cite{NARBERT}) &4.90 &5.50 \\
		\hline
		LASO with BERT (\cite{FNAR-BERT}) &5.20 &5.80 \\
		\hline
		KT-RL-ATT (\cite{CTCBERT1}) &4.38 &4.73 \\
		\hline
		Wav2vec-BERT (\cite{wav2vecBERTSLT2022}) &4.10 &4.39 \\
		\hline
		\textbf{ConformerAdpt+CTC-OT-BERT} &\textbf{3.96} &\textbf{4.27} \\		
		\hline		
	\end{tabular}
	\label{tab1}
\end{table}
In this table, the `Conformer+CTC' represents the baseline system, which was trained using CTC loss without utilization of an external PLM for linguistic information transfer. `Conformer+CTC/AED' denotes a hybrid CTC/AED ASR system \cite{Kim2017, Hori2017,Watanabe2017} which used a transformer decoder with attention to text representation during model training, and after the model is trained, only the conformer encoder module is used for speech recognition (the results are our implementation based on \cite{wenet2.0}). The other systems in comparison are all based on integrating acoustic and linguistic features for ASR. Specifically, they all employed the BERT model for linguistic knowledge transfer. In addition, some of them (e.g., `KT-RL-ATT' \cite{CTCBERT1}, and `Wav2vec-BERT' \cite{wav2vecBERTSLT2022}) even took pretrained acoustic model (from wav2vec2.0 \cite{wav2vec2.0}) and PLM for knowledge transfer. Based on the information in this table, it is evident that incorporating cross-modal linguistic knowledge transfer during model training to enable the acoustic encoder to learn rich linguistic information contributes to the improvement of ASR performance. Our proposed cross-modal knowledge transfer, based on OT, yields competitive results. Please note that different systems were implemented with distinct model architectures and training procedures.      
\subsection{Ablation study}
In our proposed model framework, two innovations are introduced, the first one is a cross-modal neural adapter (represented by gray blocks in Fig. \ref{fig:framework}), and the second one is the use of OT matching for cross-modal alignment. It is important to determine the contribution of each innovation. We conducted two additional experiments. The first involved attaching the neural adapter to the original conformer-based acoustic encoder and training the system using CTC-based loss only. In the second experiment, we trained the system with both CTC and cross-modal alignment losses, but without attaching the adapter to the original conformer-based acoustic encoder. The results are shown in Table \ref{tab2}.   
\begin{table}[tb]
	\centering
	\caption{Ablation experiments, CER (\%).}
	\begin{tabular}{|c||c||c|}		
		\hline		
		ConformerAdpt+CTC &5.64 &6.24 \\ 
		\hline		
		Conformer+CTC-OT-BERT &4.33 &4.79 \\ 
		\hline	
	\end{tabular}
	\label{tab2}
\end{table}
In this table, the entry labeled `ConformerAdpt+CTC' corresponds to the branch enclosed within the dashed red box in Fig. \ref{fig:framework}. The performance is even worse than that of the `Conformer+CTC (baseline)' (refer to table \ref{tab1}). This indicates that using only the conformer with an adapter architecture, without OT-based cross-modal transfer learning, does not yield satisfactory results. `Conformer+CTC-OT-BERT' indicates that the adapter does not connect to the conformer encoder for feature representation, despite the utilization of OT-based cross-modal transfer learning in model learning. Based on these results, we can observe that the performance is quite satisfactory, likely due to the sharing of the conformer encoder during cross-modal learning. Furthermore, by examining the results of `ConformerAdpt+CTC-OT-BERT' in table \ref{tab1}, we can observe that explicitly incorporating the adapter for linguistic feature extraction further enhances the performance.

\section{Conclusion}
\label{sec:conclusion}
In this study, we introduced a novel cross-modal alignment and knowledge transfer model for ASR leveraging the concept of OT. In the model, an OT matching module was employed to align the acoustic feature sequence (from acoustic modality) with linguistic feature sequence (from the text modality). Using the obtained alignment, a transport plan was derived to transform the latent acoustic features into the text modality space, facilitating sequence matching. Additionally, we designed a cross-modal neural adapter to integrate linguistic knowledge-guided acoustic feature with the original acoustic features for ASR. Our experimental results confirm the effectiveness of linguistic knowledge transfer in enhancing ASR performance. 

Model training involves several hyper-parameters that play crucial roles in cross-modal knowledge transfer for performance improvement. In our paper, we conducted partial investigations on their value ranges, which resulted in fairly good performance. In our future work, we plan to extensively explore the impact of these hyper-parameters on ASR performance through rigorous experimentation.   


\bibliographystyle{IEEEbib}

\begin{thebibliography}{1}
\bibitem{Li2022}
J. Li, ``Recent advances in end-to-end automatic speech recognition," \emph{APSIPA Transactions on Signal and Information Processing}, DOI 10.1561/116.00000050, 2022.

\bibitem{CTCASR}
A. Graves, and N. Jaitly, ``Towards end to-end speech recognition with recurrent neural networks," in \emph{Proc. ICML}, pp. 1764–1772, 2014.


\bibitem{Chan2016}
W. Chan, N. Jaitly, Q. Le and O. Vinyals, ``Listen, attend and spell: A neural network for large vocabulary conversational speech recognition," in \emph{Proc. of ICASSP}, pp. 4960-4964, 2016.


\bibitem{Kim2017}
S. Kim, T. Hori, and S. Watanabe, ``Joint CTC-attention based end-to-end speech recognition using multi-task learning," in \emph{Proc. of ICASSP}, pp. 4835–4839, 2017.

\bibitem{Hori2017}
T. Hori, S. Watanabe, and J. R. Hershey, ``Joint ctc/attention decoding for end-to-end speech recognition," in \emph{Proc. of ACL}, vol. 1, pp. 518–529, 2017.

\bibitem{Watanabe2017}
S. Watanabe, T. Hori, S. Kim, J. R. Hershey and T. Hayashi, ``Hybrid CTC/Attention Architecture for End-to-End Speech Recognition," \emph{IEEE Journal of Selected Topics in Signal Processing}, vol. 11, no. 8, pp. 1240-1253, 2017.

\bibitem{RNNTASR}
A. Graves, ``Sequence transduction with recurrent neural networks," \emph{arXiv preprint}, arXiv:1211.3711, 2012.

\bibitem{BERTScore}
J. Shin, Y. Lee, and K. Jung, ``Effective sentence scoring method using BERT for speech recognition," in \emph{Proc. of ACML}, pp. 1081-1093, 2019.

\bibitem{MLMScore}
J. Salazar, D. Liang, T. Nguyen, K. Kirchhoff, ``Masked Language Model Scoring," in \emph{Proc. of ACL}, pp. 2699-2712, 2020.

\bibitem{T5}
J. Ao, R. Wang, Z. Zhou, et al, ``Speecht5: Unified-modal encoder-decoder pre-training for spoken language processing," \emph{arXiv preprint}, arXiv:2110.07205, 2021.

\bibitem{SXLSR2022}
S. Khurana, A. Laurent, J. Glass,``SAMU-XLSR: Semantically-Aligned Multimodal Utterance-Level Cross-Lingual Speech Representation," \emph{IEEE J. Sel. Top. Signal Process.}, 16(6), pp. 1493-1504, 2022.

\bibitem{HierarchicalCTC}
Higuchi, K. Karube, T. Ogawa, et al., ``Hierarchical conditional end-to-end asr with ctc and multi-granular subword units," in \emph{Proc. of ICASSP}, pp. 7797-7801, 2022.

\bibitem{intermediateCTC}
Y. Fujita, T. Komatsu, and Y. Kida, ``Multi-sequence intermediate conditioning for ctc-based asr," \emph{arXiv preprint}, arXiv:2204.00175, 2022.

\bibitem{wav2vec2.0}
A. Baevski, Y. Zhou, A. Mohamed, and M. Auli, ``Wav2vec 2.0: A framework for self-supervised learning of speech representations," in \emph{Proc. of NeurIPS}, 2020. 

\bibitem{BERT}
J. Devlin, M. Chang, K. Lee, and K. Toutanova, ``Bert: Pretraining of deep bidirectional transformers for language understanding," \emph{arXiv preprint}, arXiv:1810.04805, 2018.	

\bibitem{CIFBERT1}
M. Han, F. Chen, J. Shi, S. Xu, B. Xu, ``Knowledge Transfer from Pre-trained Language Models to Cif-based Speech Recognizers via Hierarchical Distillation," \emph{arXiv preprint}, arXiv:2301.13003, 2023.

\bibitem{CTCBERT1}
K. Deng, S. Cao, Y. Zhang, L. Ma, G. Cheng, J. Xu, P. Zhang, ``Improving CTC-Based Speech Recognition Via Knowledge Transferring from Pre-Trained Language Models," in \emph{Proc. of ICASSP}, pp. 8517-8521, 2022.

\bibitem{CTCBERT2}
K. Deng, Z. Yang, S. Watanabe, Y. Higuchi, G. Cheng, P. Zhang, ``Improving Non-Autoregressive End-to-End Speech Recognition with Pre-Trained Acoustic and Language Models," in \emph{Proc. of ICASSP}, pp. 8522-8526, 2022.

\bibitem{CIFBERT2}
M. Han, L. Dong, Z. Liang, M. Cai, S. Zhou, Z. Ma, B. Xu, ``Improving End-to-End Contextual Speech Recognition with Fine-Grained Contextual Knowledge Selection," in \emph{Proc. of ICASSP}, pp. 8532-8536, 2022

\bibitem{Cho2020}
W. Cho, D. Kwak, J. Yoon, N. Kim, ``Speech to Text Adaptation: Towards an Efficient Cross-Modal Distillation," in \emph{Proc. of INTERSPEECH}, pp. 896-900, 2020.

\bibitem{FNAR-BERT}
Y. Bai, J. Yi, J. Tao, Z. Tian, Z. Wen and S. Zhang, "Fast End-to-End Speech Recognition Via Non-Autoregressive Models and Cross-Modal Knowledge Transferring From BERT," \emph{IEEE/ACM Transactions on Audio, Speech, and Language Processing}, vol. 29, pp. 1897-1911, 2021.

\bibitem{wav2vecBERTSLT2022}
K. Lu and K. Chen, ``A Context-aware Knowledge Transferring Strategy for CTC-based ASR," in \emph{Proc. of SLT}, pp. 60-67, 2022.

\bibitem{NARBERT}
F. Yu, K. Chen, and K. Lu, ``Non-autoregressive ASR Modeling using Pre-trained Language Models for Chinese Speech Recognition," \emph{IEEE/ACM Transactions on Audio, Speech, and Language Processing}, vol. 30, pp. 1474-1482, 2022



\bibitem{KD2015}
G. Hinton, O. Vinyals, and J. Dean, ``Distilling the knowledge in a neural network," \emph{arXiv preprint}, arXiv:1503.02531, 2015.

\bibitem{Choi2022}
K. Choi, H. Park, ``Distilling a Pretrained Language Model to a Multilingual ASR Model," in \emph{Proc. of INTERSPEECH}, pp. 2203-2207, 2022.

\bibitem{Cross2021}
W. Wang, S. Ren, Y. Qian, S. Liu, Y. Shi, Y. Qian, M. Zeng, ``Optimizing Alignment of Speech and Language Latent Spaces for End-To-End Speech Recognition and Understanding," in \emph{Proc. of ICASSP}, pp. 7802-7806, 2021.

\bibitem{Futami2022}
H. Futami, H. Inaguma, M. Mimura, S. Sakai, T. Kawahara, ``Distilling the Knowledge of BERT for CTC-based ASR," \emph{arXiv preprint}, CoRR abs/2209.02030, 2022. 

\bibitem{Higuchi2023}
Y. Higuchi, T. Ogawa, T. Kobayashi, S. Watanabe, ``BECTRA: Transducer-Based End-To-End ASR with Bert-Enhanced Encoder," in \emph{Proc. of ICASSP}, pp. 1-5, 2023.

\bibitem{Transformer}
A. Vaswani, N. Shazeer, N. Parmar, J. Uszkoreit, L. Jones, A. Gomez, L. Kaiser, and I. Polosukhin, ``Attention is all you need," in \emph{Proc. of NIPS}, pp. 5998-6008, 2017.		

\bibitem{VillanoBook}
C. Villani, Optimal transport: old and new, volume 338. Springer, 2009

\bibitem{CourtyNIPS2017}
N. Courty, R.  Flamary, A. Habrard, A. Rakotomamonjy, ``Joint distribution optimal transportation for domain adaptation," in \emph{Proc. of NIPS}, pp. 3733-3742, 2017.

\bibitem{ICLR2023}
H. Tseng, H. Lin, H. Hsuan, and Y. Tsao, ``Interpretations of Domain Adaptations via Layer Variational Analysis," \emph{arXiv preprint}, CoRR abs/2302.01798, 2023.

\bibitem{LuICASSP2021}
X. Lu, P. Shen, Y. Tsao, H. Kawai, ``Unsupervised Neural Adaptation Model Based on Optimal Transport for Spoken Language Identification," in \emph{Proc. of ICASSP}, pp. 7213-7217, 2021.

\bibitem{Lin2021}
H. Lin, H. Tseng, X. Lu, Y. Tsao, ``Unsupervised Noise Adaptive Speech Enhancement by Discriminator-Constrained Optimal Transport," in \emph{Proc. of NeurIPS}, pp. 19935-19946, 2021.


\bibitem{ACL2023}
Y. Zhou, Q. Fang, Y. Feng, ``CMOT: Cross-modal Mixup via Optimal Transport for Speech Translation," \emph{arXiv preprint},  arXiv:2305.14635, 2023.

\bibitem{ICML2023}
P. Le, H. Gong, C. Wang, J. Pino, B. Lecouteux, D. Schwab, ``Pre-training for Speech Translation: CTC Meets Optimal Transport," \emph{arXiv preprint}, CoRR abs/2301.11716, 2023.

\bibitem{conformer2020}
A. Gulati, J. Qin, C. Chiu, et al., ``Conformer: Convolution augmented transformer for speech recognition," \emph{arXiv preprint}, arXiv:2005.08100, 2020

\bibitem{Chi2021}
Z. Chi, L. Dong, B. Zheng, S. Huang, X. Mao, H. Huang, and F. Wei, ``Improving Pretrained Cross-Lingual Language Models via Self-Labeled Word Alignment," in \emph{Proc. of ACL}, pp. 3418-3430, 2021.

\bibitem{OPT}
M. Cuturi, “Sinkhorn distances: Lightspeed computation of optimal transport,” in \emph{Proc. of NIPS}, vol. 26, 2013.

\bibitem{AISHELL1} 
Hui Bu, Jiayu Du, Xingyu Na, Bengu Wu, and Hao Zheng, ``AIShell-1: An open-source mandarin speech corpus and a speech recognition baseline,” in \emph{Proc. of COCOSDA}, pp. 1-5, 2017.

\bibitem{Adam}
Diederik P. Kingma, Jimmy Ba, ``Adam: A Method for Stochastic Optimization," in \emph{Proc. of ICLR}, 2015.

\bibitem{Huggingface}
https://huggingface.co/

\bibitem{wenet2.0}
B. Zhang, D. Wu, Z. Peng, X. Song, Z. Yao, H. Lv, L. Xie, C. Yang, F. Pan, J. Niu, ``WeNet 2.0: More Productive End-to-End Speech Recognition Toolkit," in \emph{Proc. of INTERSPEECH}, pp. 1661-1665, 2022.


\end{thebibliography}

\end{document}